\documentclass{aa}
\usepackage[varg]{txfonts}
\usepackage{natbib}
\usepackage{graphicx}
\usepackage{mathrsfs}

\newcommand{\CII}{\ion{C}{II}\ }

\usepackage{color}

\begin{document}

\title{\textit{Herschel} HIFI observations of ionised carbon in the $\beta$~Pictoris debris disk \thanks{\emph{Herschel} is an ESA space observatory with science instruments provided by European-led Principal Investigator consortia and with important participation from NASA.}}
\author{G. Cataldi\inst{1,2} \and A. Brandeker\inst{1,2} \and G. Olofsson\inst{1,2} \and B. Larsson\inst{1} \and R. Liseau\inst{3} \and J. Blommaert\inst{4} \and M. Fridlund\inst{5,6} \and R. Ivison\inst{7,8} \and E. Pantin\inst{9} \and B. Sibthorpe\inst{10} \and B. Vandenbussche\inst{4} \and Y. Wu\inst{11}}
\institute{AlbaNova University Centre, Stockholm University, Department of Astronomy, SE-106 91 Stockholm, Sweden \and
Stockholm University Astrobiology Centre, SE-106 91 Stockholm, Sweden \and 
Department of Earth and Space Sciences, Chalmers University of Technology, Onsala Space Observatory, Onsala, Sweden \and
Instituut voor Sterrenkunde, KU Leuven, Celestijnenlaan 200D, 3001 Leuven, Belgium \and
Institute of Planetary Research, German Aerospace Center, Rutherfordstrasse 2, 124 89 Berlin, Germany \and
Leiden Observatory, University of Leiden, P.O. Box 9513, NL-2300 RA Leiden, The Netherlands \and
UK Astronomy Technology Centre, Royal Observatory, Blackford Hill, Edinburgh EH9 3HJ, United Kingdom \and
Institute for Astronomy, University of Edinburgh, Blackford Hill, Edinburgh EH9 3HJ, United Kingdom \and
Laboratoire AIM, CEA/DSM - CNRS - Universit\'e Paris Diderot, IRFU/Service d'Astrophysique, B\^at.\,709, CEA-Saclay, 91191 Gif-sur-Yvette Cedex, France \and
SRON Netherlands Institute for Space Research, Landleven 12, 9747 AD Groningen, The Netherlands \and
Department of Astronomy and Astrophysics, University of Toronto, ON M5S 3H4, Canada}

\date{Received ; accepted}

\abstract
 % context heading (optional), leave it empty if necessary
{The dusty debris disk around the $\sim$20\,Myr old main-sequence A-star $\beta$~Pictoris is known to contain gas. Evidence points towards a secondary origin of the gas as opposed to being a direct remnant from the initial protoplanetary disk, although the dominant gas production mechanism is so far not identified. The origin of the observed overabundance of C and O compared with solar abundances of metallic elements such as\ Na and Fe is also unclear.}
 % aims heading (mandatory)
{Our goal is to constrain the spatial distribution of C in the disk, and thereby the gas origin and its abundance pattern.}
 % methods heading (mandatory)
{We used the HIFI instrument onboard the \textit{Herschel Space Observatory} to observe and spectrally resolve \CII emission at 158\,$\mu$m from the $\beta$~Pic debris disk. Assuming a disk in Keplerian rotation and a model for the line emission from the disk, we used the \textit{spectrally} resolved line profile to constrain the \textit{spatial} distribution of the gas.}
 % results heading (mandatory)
 {We detect the \CII 158\,$\mu$m emission. Modelling the shape of the emission line shows that most of the gas is located at
about $\sim$100\,AU or beyond. We estimate a total C gas mass of $1.3_{-0.5}^{+1.3}\times10^{-2}$\,M$_\oplus$ (central 90\% confidence interval). The data suggest that more gas is located on the southwest side of the disk than on the northeast side. The shape of the emission line is consistent with the hypothesis of a well-mixed gas (constant C/Fe ratio throughout the disk). Assuming instead a spatial profile expected from a simplified accretion disk model, we found it to give a significantly poorer fit to the observations.}
 % conclusions heading (optional)
{Since the bulk of the gas is found outside 30\,AU, we argue that the cometary objects known as ``falling evaporating bodies'' are probably not the dominant source of gas; production from grain-grain collisions or photodesorption seems more likely. The incompatibility of the observations with a simplified accretion disk model might favour a preferential depletion explanation for the overabundance of C and O, although it is unclear how much this conclusion is affected by the simplifications made. More stringent constraints on the spatial distribution will be available from ALMA observations of \ion{C}{I} emission at 609\,$\mu$m.}
   

\maketitle

%%%%%%%%%%%%%%%%%%%%%%%%%%%%%%%%%%%%%%%%%%%%%%%%%%%%%%%%%%%%%%%%%%%%%%%

\section{Introduction}
Gas-rich and dense protoplanetary disks around young stars are thought to be the birthplaces of planetary systems. Planet formation occurs together with various processes that gradually clear the disk from gas on a timescale of $\sim$10~Myr \citep[see, e.g.,][]{Haisch_etal_2001,Jayawardhana_etal_2006,Hillenbrand_2008,Mamajek_2009}. The removal of the gas results in an optically thin debris disk consisting mostly of dust and the planetesimals remaining from the protoplanetary phase. These objects can be seen as analogues to the Kuiper belt or the asteroid belt in the solar system. Although the formation of gas giants ceases in debris disks because of the lack of gas, models suggest that terrestrial planet formation may still be ongoing for several 100\,Myr \citep{Kenyon_etal_2006}. In addition, planet migration can still change the architecture of the system.

Debris disks are typically detected by an excess in infrared emission above the photosphere of the hosting star. \citet{Aumann_etal_1984} were the first to report excess emission observed with IRAS around $\alpha$ Lyrae (Vega), interpreted as thermal emission from solid particles around the star. Later, this ``Vega-phenomenon'' was also discovered for $\beta$~Pictoris and other stars \citep{Aumann_1985}, with the first image of a disk observed by \citet{Smith_etal_1984} around $\beta$~Pic. Today, it is estimated that $\sim$15\% of all stars are surrounded by a debris disk \citep{Wyatt_2008}. The dust particles in debris disks are blown out of the system by radiation pressure or fall onto the star by Poynting-Robertson drag on timescales much shorter than the lifetime of the system. Therefore one needs a mechanism to replenish the grain concentration, most likely collisions of larger bodies \citep{Backman_etal_1993}. The observed quantity of dust in a system can thus give a clue about the total mass residing in larger solid bodies (planetesimals or comets not incorporated into planets).

A few debris disks show observable amounts of gas in addition
to the dust. Examples of gaseous debris disks include HD\,172555 \citep[see e.g.][]{Riviere-Marichalar_etal_2012}, HD\,32297 \citep{Redfield_2007,Donaldson_etal_2013}, HD\,21997 \citep{Moor_etal_2011}, 49~Ceti \citep{Roberge_etal_2013}, and $\beta$~Pic, the object of interest in this work. Atomic carbon has so far only been detected around a few systems, including $\beta$~Pic, 49~Cet and HD\,32297. The unexpected presence of gas at these late stages of disk evolution makes such systems particularly interesting. The gas could either be primordial, that is,\ a remnant from the protoplanetary phase, or be of secondary origin.

In this work, we present observations of the gaseous debris disk around $\beta$~Pic, a young (10--20~Myr, \citet{Zuckerman_etal_2001,Binks_Jeffries_2014}) A6V star \citep{Gray_etal_2006} at a distance of $19.44\pm0.05$~pc \citep{vanLeeuwen_2007}. $\beta$~Pic hosts an outstanding debris disk that has been subject to intensive studies during the past decades. The discovery of a giant planet ($\beta$~Pic~b) on an orbit of 10\,AU by \citet{Lagrange_etal_2010} further increased the interest in the system. Observations of sharp absorption lines suggested the presence of circumstellar gas early on \citep{Slettebak_1975,Slettebak_etal_1983,Kondo_etal_1985,Hobbs_etal_1985} and led to the classification of $\beta$~Pic as a shell star. More recent observations showed that the gas is spatially extended in the disk and in Keplerian rotation \citep{Olofsson_etal_2001,Brandeker_etal_2004}. The gas is probably not a direct remnant from the protoplanetary phase. Rather, it is thought to be of secondary origin, that
is,\ steadily produced in the disk, probably from the dust itself \citep{Fernandez_etal_2006}. Different gas-producing mechanisms such as falling evaporating bodies (FEBs, planetesimals evaporating in immediate vicinity to the star, \citet{Beust_etal_2007}), photon-stimulated desorption of dust \citep{Chen_etal_2007}, or collisional vaporisation of dust \citep{Czechowski_etal_2007} have been proposed as the origin of the gas. Colliding comets can potentially produce large amounts of CO \citep{Zuckerman_Song_2012}, which in turn could be the origin of the C and O gas in the disk. If the gas production is indeed linked to the destruction of dusty or cometary material, one could study the dust composition by observing the gas derived from it. This would require a precise understanding of the gas-production mechanisms, however, and it is one of the aims of this work to help distinguish between different possible production scenarios.

The composition of $\beta$~Pic's circumstellar gas is quite varied. \citet{Brandeker_etal_2004} reported the detection of Na, Fe, Ca, Cr, Ti, and Ni in emission and resolved their spatial distribution in the disk. Given that radiation pressure largely dominates gravitation for some of the observed species, the question arose why the gas is seen in Keplerian rotation around the star and not blown out. \citet{Fernandez_etal_2006} noticed that species affected by radiation pressure are ionised and concluded that Coulomb interactions could brake the gas, provided that carbon is overabundant. Observations by \citet{Roberge_etal_2006} showed that the gas is indeed enhanced in carbon. Subsequent absorption data of Fe and Na obtained by HARPS at the ESO 3.6m telescope \citep{Brandeker_2011} and \textit{Herschel}/PACS observations of the \CII emission at 158\,$\mu$m \citep{Brandeker_etal_2014_subm} show that the carbon overabundance is even higher than previously inferred, providing thus an explanation for the observed gas dynamics, but posing the question about the origin of such a carbon-rich gas disk. \citet{Xie_etal_2013} theoretically studied preferential production of carbon (and oxygen) and preferential depletion of the other elements as a possible mechanism causing the carbon overabundance. They concluded that information about the spatial distribution of the carbon gas would be useful to distinguish between these possibilities. In contrast to other species that have been spatially mapped \citep[see][]{Brandeker_etal_2004,Nilsson_etal_2012}, the spatial profile of the \CII gas is unconstrained, although carbon is a dominant component by mass. With current instrumentation it is indeed not possible to resolve the \CII emission. In this work, we thus opted for an alternative approach and tried to constrain the spatial profile using spatially unresolved, but spectrally resolved data. We observed the $\beta$~Pic disk in \CII 158\,$\mu$m emission with \textit{Herschel}/HIFI. This line was already detected with PACS \citep{Brandeker_etal_2014_subm}, but only HIFI is able to spectrally resolve it, thus allowing us to extract spatial information. By giving clues about the location of the carbon gas, our work is also expected to help constrain the C/O ratio, a fundamental parameter for planet formation. Indeed, conversion of an observed O flux into an O mass is only possible when the excitation of the gas can be estimated. This requires knowledge of the location of the O gas in the disk. Assuming that C and O are well mixed, constraints on the location of C can be used to derive an O mass from spatially unresolved observations of O like those by \citet{Brandeker_etal_2014_subm}.

This paper is organised as follows: we describe the data reduction in Sect.~\ref{sec:obs_data_reduction} and the results of the observations in Sect.~\ref{sec:results}. We then test whether the data are consistent with a well-mixed gas or an accretion profile, constrain the density profile of the carbon gas and assess a possible asymmetry of the disk (Sect.~\ref{sec:analysis}). In Sect.~\ref{sec:discussion}, we discuss the implications on the nature of the gas-producing mechanism, compare our data with previous observations of \CII in the $\beta$~Pic disk and with gaseous debris disks around A-type stars in general. In addition, we present a prediction for the detectability of \ion{C}{I} emission at 609\,$\mu$m with ALMA. Our results are summarised in Sect.~\ref{sec:summary}.

\section{Observations and data reduction}\label{sec:obs_data_reduction}
\begin{figure}
\resizebox{\hsize}{!}{\includegraphics{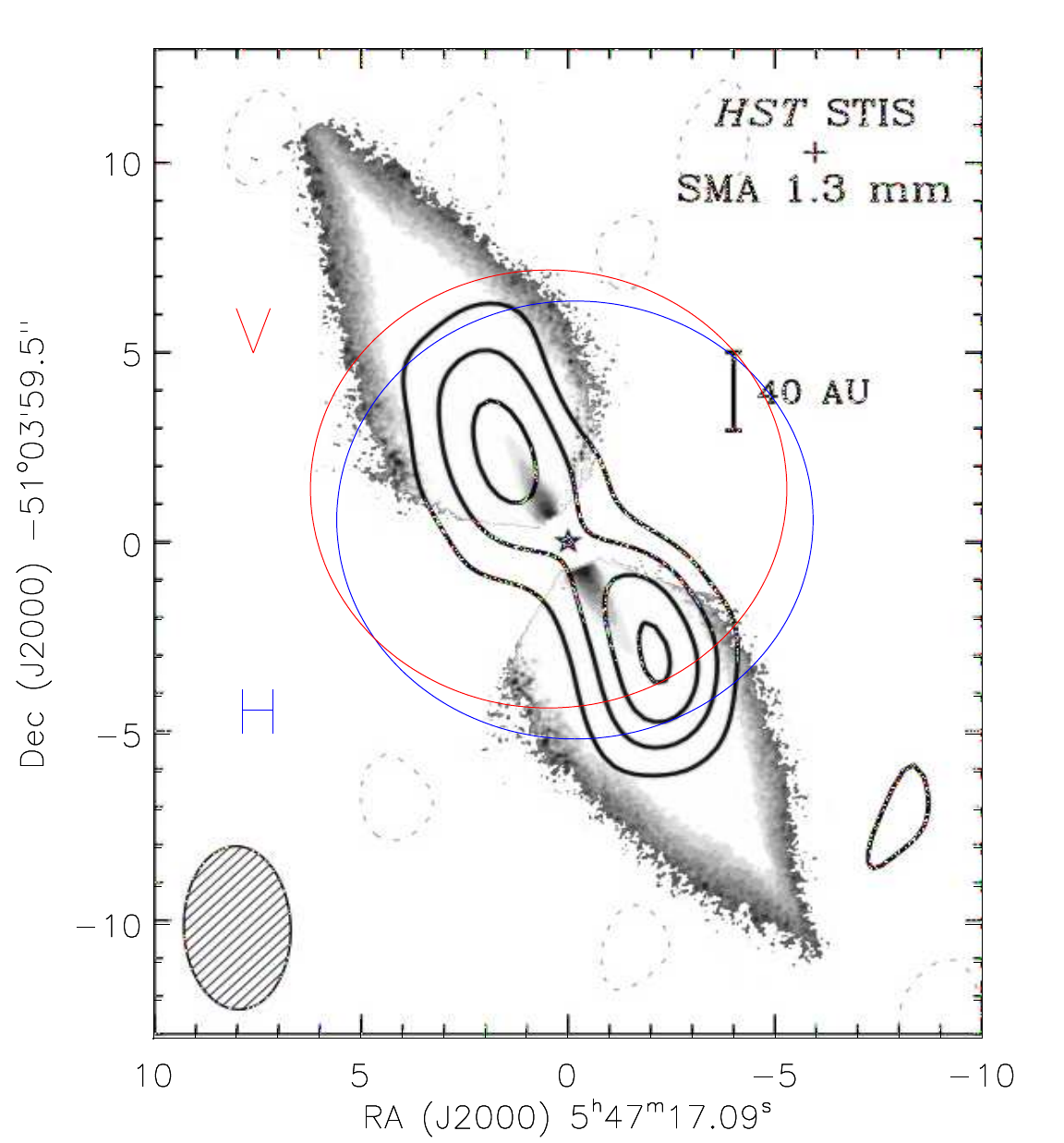}}
\caption{Pointing of H- and V-polarisation beams. The beam centres are 1.3$\arcsec$ apart. The diameter of the circles corresponds to the HPBW. The background image of the $\beta$~Pic disk shows contours of the millimetre emission and a grey scale of optical scattered light \citep{Wilner_etal_2011}.}
\label{HIFI_beam_pointing}
\end{figure}

We used the Heterodyne Instrument for the Far Infrared \citep[HIFI,][]{deGraauw_etal_2010,Roelfsema_etal_2012} onboard the \textit{Herschel Space Observatory} \citep{Pilbratt_etal_2010} to observe \CII 158\,$\mu$m emission from the $\beta$~Pic disk. The observations were carried out on operating day OD 985 (2012 January 24; observation ID 1342238190) in dual beam-switch mode (nod of 3\arcmin) with continuum optimisation and fast chopping to reduce the effects of standing waves as much as possible. We used the Wide Band Spectrometer (WBS) with a spectral resolution of 0.17\,km\,s$^{-1}$ for HIFI band 7, covering the wavelength range 157--176 $\mu$m. The total observing time was 8.33~h. The horizontal (H) and vertical (V) polarisation beams were pointed at slightly different locations (distance between the beam centres 1.3\arcsec, see Fig.~\ref{HIFI_beam_pointing}), with the idea of potentially extracting additional spatial information. The data were calibrated in a standard way using the Herschel Interactive Processing Environment (HIPE) version 11 \citep{Ott_2010}. A main-beam efficiency of 0.69 was adopted. An additional reduction step included the removal of ripples in the data. For the analysis and fitting, we oversampled the data and chose a bin size of 0.08\,km\,s$^{-1}$. The oversampling is the reason for the apparent correlations in the data, discussed in Appendix~\ref{details_error_analysis}. Fig.~\ref{fig:HIFI_data_plots} shows the calibrated spectra in units of main-beam brightness temperature $T_\mathrm{mb}$. For better visualisation, the data have been plotted with a bin width of 0.63\,km\,s$^{-1}$. The half power beam width (HPBW) at the wavelength of the \CII emission line is $\sim$11$\arcsec$ \citep{Roelfsema_etal_2012}, corresponding to $\sim$200\,AU at the distance of $\beta$~Pic.

\begin{figure*}
\includegraphics[width=16.4cm]{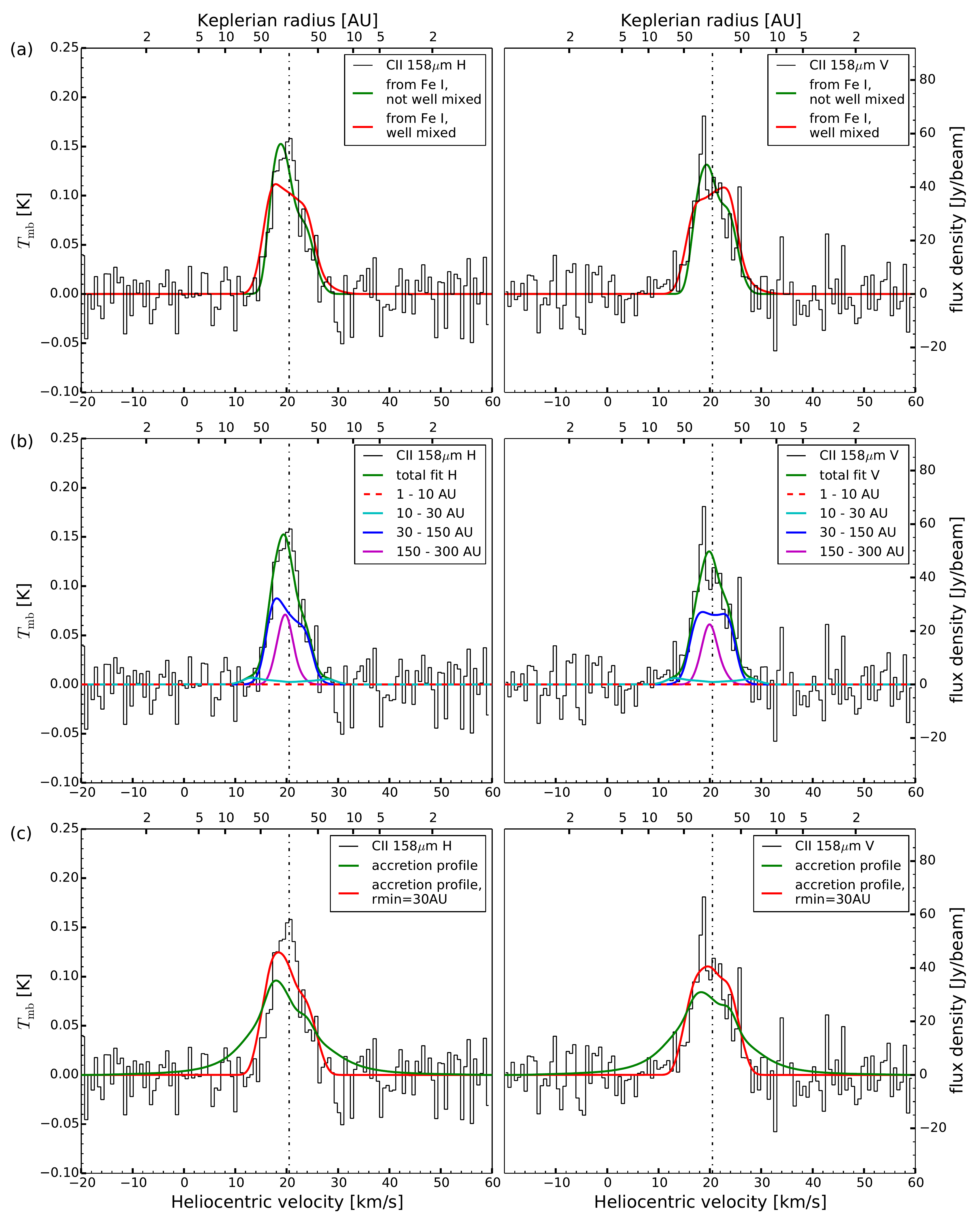}
\caption{
\textit{Herschel}/HIFI spectra of the \CII 158\,$\mu$m emission from the $\beta$~Pic debris disk and different fits. The two columns show horizontal (H) and vertical (V) polarisation. The Keplerian radius $r$ (related to the velocity axis by $v=(GM_*r^{-1})^{0.5}$) is plotted on the upper x-axis of each data set. It corresponds to the largest radius where the corresponding projected velocity on the lower x-axis can be reached. If there is gas inside a certain Keplerian radius, we should see emission at the corresponding velocity (assuming a non-clumpy disk). The vertical dashed line indicates $\beta$~Pic's systematic velocity of 20.5\,km\,s$^{-1}$ \citep{Brandeker_2011}. For better visualisation the data have been rebinned to a bin size of 0.63\,km\,s$^{-1}$. The original data have a resolution of 0.17\,km\,s$^{-1}$. For all fits, we oversampled the data to bin sizes of 0.08\,km\,s$^{-1}$. Note that the difference between the fitted H and V beam profiles comes from the different positioning of the beams (Fig.~\ref{HIFI_beam_pointing}).
Panel \textbf{(a)} compares the data with \CII profiles derived from observations of \ion{Fe}{I} by \citet{Nilsson_etal_2012}. Assuming a well-mixed gas, a reasonable fit can be found. If the assumption is dropped, the fit can be improved. See the text (Sect.~\ref{subsec:well_mixed_gas}) for more details.
Panel \textbf{(b)} shows the best ring fit profile and the contribution of individual rings. Emission coming from the two inner rings is minor.
Panel \textbf{(c)} shows the fitted profile for an accretion surface density, $\Sigma\propto r^{-1}$. It is difficult to bring this model into agreement with the observations, because the gas close to the star produces extended wings in the profile. A profile with the same underlying surface density, but truncated at 30\,AU, gives a significantly better fit and demonstrates the problem.
}
\label{fig:HIFI_data_plots}
\end{figure*}

\section{Results}\label{sec:results}
Table~\ref{HIFI_CII_emission} shows the \CII line emission strengths detected by \textit{Herschel}/HIFI. Statistical errors on the emission were calculated in the following manner: we first fitted a Gaussian to the line and determined its full width at half maximum (FWHM). The data were then rebinned with a bin size corresponding to twice the measured FWHM. The bins were arranged such that the peak of the fitted Gaussian fell on the centre of one of the bins. The flux stated in Table~\ref{HIFI_CII_emission} is the flux contained in this bin; the stated statistical errors were calculated as the root mean square (rms) of the fluxes contained in the bins outside the line. Note that the \textit{flux calibration} uncertainty of HIFI is $\sim$10\% \citep{Roelfsema_etal_2012}.

Although the two polarisation beams were pointed at slightly different locations (Fig.~\ref{HIFI_beam_pointing}), the two resulting spectra are statistically indistinguishable. We nevertheless took the different beam positions into account in our analysis. The effect of the different beam positions shows up more clearly when calculating model profiles (Fig.~\ref{fig:HIFI_data_plots}).

Visual inspection of the data in Fig.~\ref{fig:HIFI_data_plots} already suggests an asymmetric distribution of \CII in the disk. There is slightly more flux for velocities lower than the systematic velocity of $\beta$~Pic, indicating emission preferentially from the SW of the disk (the sense of rotation was determined by \citet{Olofsson_etal_2001}; the disk rotates towards us in the SW). We assess the significance of this feature in section \ref{subsec:asymmNESW}.

\begin{table*}
\caption{HIFI detected \CII 158\,$\mu$m emission from the $\beta$~Pic disk. Quoted uncertainties represent 1$\sigma$ statistical errors estimated as described in the text (Sect.~\ref{sec:results}). The flux calibration uncertainty of HIFI is $\sim$10\% \citep{Roelfsema_etal_2012}.}
\label{HIFI_CII_emission}
\centering % used for centering table
\begin{tabular}{c c c} % centered columns
\hline\hline % inserts double horizontal lines
Beam & $\int T_{\mathrm{mb}}\,\mathrm{d}v$  & Flux/beam \\ % table heading
 & (K km\,s$^{-1}$) & (erg\,s$^{-1}$\,cm$^{-2}$\,beam$^{-1}$) \\
\hline % inserts single horizontal line
H & $1.01\pm0.02$  & $(2.36\pm0.06) \times 10^{-14}$\\ % inserting body of the table
V & $1.07\pm0.02$ & $(2.50\pm0.05) \times 10^{-14}$\\
\hline
\end{tabular}
\end{table*}

\section{Analysis}\label{sec:analysis}
The aim of the this work is to set constraints on the radial density profile of C using the emission spectrum from HIFI. To model the disk, we assumed perfect edge-on geometry and Keplerian rotation of the gas. Although the velocity pattern of \CII has never been measured directly, Keplerian motion has been observed for other species such as \ion{Na}{I} \citep{Olofsson_etal_2001,Brandeker_etal_2004}. Moreover, \CII is not affected by radiation pressure, making Keplerian rotation a reasonable hypothesis. We wrote a code that, given a hypothetical radial gas-density profile as input, calculates the resulting HIFI spectrum. To do so, we first invoked the \textsc{ontario} code to calculate the \CII emission in every point of the disk. \textsc{ontario} was specifically designed to address gas emission from debris disks around A and F stars \citep{Zagorovsky_etal_2010}. It calculates the gas ionisation and thermal balance self-consistently, and computes the level populations of species of interest using statistical equilibrium. For the models we consider in this work, the calculated gas temperature peaks at $\sim$60\,K in the region where most of the gas is located (around 100\,AU). The end product of \textsc{ontario} is a two-dimensional grid (radius and height in disk) of line luminosity densities. The second step is then to project the line luminosities along the line of sight, taking the rotation into account. Finally the projected map is multiplied with the HIFI beam, and integrated to provide synthetic observations of the disk.

From initial estimates of the \CII spatial distribution, we found that the 158\,$\mu$m mid-plane optical depth was of the order of 1, violating the assumption of optically thin emission of \textsc{ontario}. This prompted us to extend \textsc{ontario} to take into account radiative transfer effects, as described in more detail by \citet{Brandeker_etal_2014_subm}. In summary, we computed the level populations of the innermost grid points first, followed by the grid points second-closest to the star, going outwards. The intensity from the star seen at each grid point is adjusted by the line extinction towards the star. As a second-order correction, the intensity from photons scattered locally was taken into account by using a photon escape formulation, following Appendix~B of \citet{Tielens_etal_1985}. The escape probability of photons was computed iteratively, starting with an assumption of complete escape (first iteration), and then using the computed levels from the first iteration to approximate the photon escape probability from the average of four directions (in, out, up, down) as described in \citet{Gorti_et_al_2004}. Since the vertical optical depth of the disk in the dominant cooling line \CII 158\,$\mu$m generally is $\lesssim$0.1, and the excitation is mainly due to colliding electrons, the change in luminosity density distribution compared with the optically thin approximation is minor ($\lesssim$10\,\%).

After determining the luminosity density distribution of the disk, we integrated the wavelength-dependent intensity along the line of sight, taking into account Doppler shifts from the Keplerian velocity field and assuming a two-level system for the \CII 158\,$\mu$m line, where the levels are also known from the \textsc{ontario} output. In this case, the optical depth is given by \citep[e.g.][]{Rybicki_Lightman_2004}
\begin{equation}\label{eq:opt_depth}
\tau_\nu=\frac{c^2}{8\pi\nu^2}A_{21}\phi_\nu\left(\frac{g_2}{g_1}N_1-N_2\right)
,\end{equation}
with $A_{21}=2.29\cdot10^{-6}$\,$s^{-1}$ the Einstein coefficient for spontaneous emission, $\phi_\nu$ the line profile (assumed to be Gaussian with a broadening parameter $b=1.5$\,km\,s$^{-1}$ and a central frequency $\nu_0=1900.7$\,GHz), and $g_2=4$ and $g_1=2$ the statistical weights of the upper and the lower level, respectively (numerical values from the NIST database \citep{NIST_ASD}). $N_2$ and $N_1$ are the column densities of ions in the upper and lower state, respectively. We implemented a standard ray-tracing method to determine the extinction along the line of sight in the following way: we divided the disk into slices perpendicular to the line of sight and considered the slice closest to the observer first. In addition to the emission from the first slice, we also computed its optical depth $\tau_{\nu,1}$ using equation \ref{eq:opt_depth}. We then computed the emission from the second slice, adjusted it by a factor $\exp(-\tau_{\nu,1})$, and computed its optical depth $\tau_{\nu,2}$. The emission of the third slice is then extinct by a factor $\exp(-\tau_{\nu,1}-\tau_{\nu,2})$, and so on until the end of the disk at the far side of the star is reached.

Table~\ref{general_model_params} shows the values of general parameters used in our disk model: the mass of $\beta$~Pic, its distance, the position angle of the disk, $\beta$~Pic's systematic velocity, and the broadening parameter $b$ of the intrinsic line profile, assumed to be Gaussian. The value for $b=1.5$\,km\,s$^{-1}$ was adapted in line with \citet{Nilsson_etal_2012}, and thus is slightly lower than the measurement by \citet{Crawford_etal_1994}. The value we used for the opening angle of the disk, $\alpha=0.2$, is close to the one adapted by \citet{Brandeker_etal_2004}. We also corrected for the proper motion of $\beta$~Pic as measured by \citet{vanLeeuwen_2007}.

\begin{table*}
\caption{Fixed disk model parameters adapted for fitting the HIFI data.} % title of Table
\label{general_model_params}
\centering % used for centering table
\begin{tabular}{c c c c} % centered columns
\hline\hline % inserts double horizontal lines
Parameter & Value & Unit & Reference \\ % table heading
\hline % inserts single horizontal line
stellar mass $M_*$& 1.75& $M_\sun$ & (1) \\ % inserting body of the table
distance $d$ & 19.44 & pc & (2) \\
position angle PA&  29.2 & $\degr$  & (3) \\
barycentric systematic velocity $v_*$ & 20.5 & km\,s$^{-1}$ & (4) \\
broadening parameter $b$ &  1.5 & km\,s$^{-1}$ & (5), (6) \\
opening angle $\alpha$&  0.2 & rad & (6), (7)\\
\hline
\end{tabular}
\tablebib{
(1)~\citet{Crifo_etal_1997}; (2)~\citet{vanLeeuwen_2007}; (3)~\citet{Lagrange_etal_2012}; (4)~\citet{Brandeker_2011}; (5)~\citet{Crawford_etal_1994}; (6)~\citet{Nilsson_etal_2012}; (7)~\citet{Brandeker_etal_2004}.
}
\end{table*}

\subsection{Is the observed line profile consistent with a well-mixed gas?}\label{subsec:well_mixed_gas}
We tested whether our observations are consistent with the hypothesis of a well-mixed gas. Spatially resolved observations of \ion{Fe}{I} by \citet{Nilsson_etal_2012} and \textsc{ontario} were used to test this assumption by finding the spatial distribution of Fe and the C/Fe ratio that best reproduced both the observed \ion{Fe}{I} spatial profile and the observed \CII spectrally resolved line. This was done iteratively by first fixing a C/Fe ratio and computing the gas distribution that precisely corresponds to the \ion{Fe}{I} profile by \citet{Nilsson_etal_2012}, and by then projecting the \CII emission for comparison with the observed spectral profile. The C/Fe ratio was then adjusted until a best fit (in a $\chi^2$ sense) was achieved. We found the best ratio to be C/Fe=300. The fit is shown in Fig.~\ref{fig:HIFI_data_plots}a and is consistent with the profile within the noise of the observations.

By dropping the assumption of a well-mixed gas, a better fit can be achieved (Fig.~\ref{fig:HIFI_data_plots}a). In this case, the C/Fe ratio can be different for the north-east (NE) and the south-west (SW). In addition, the C gas disk was allowed to be truncated inside at a certain radius even when \ion{Fe}{I} was known to extend farther inward. The introduction of an inner hole reduces the emission in the wings of the profile, which seem too broad for a well-mixed gas. Under this conditions we found best-fit C/Fe ratios of 225 (NE) and 2900 (SW) and truncation radii of 30\,AU (NE) and 97\,AU (SW). The improvement of the fit is marginal, but is indicative of a possible elemental separation between C and Fe, where proportionally more C is located in the SW and at larger radii.

\subsection{Is the observed line profile consistent with an accretion disk?}
The origin of the gas in the $\beta$~Pic debris disk is still unclear. \citet{Xie_etal_2013} studied a model of the temporal evolution in elemental composition of the gas to explain the extreme overabundance of carbon and oxygen compared with metallic elements, for example,\ Na and Fe. Two competing scenarios were considered: preferential production (the gas is produced at the unusual composition) and preferential depletion (the gas is produced at normal abundances, but some elements are preferentially removed from the system). For the first case to be dominant, accretion of the gas has to be stronger than the differential separation of elements due to radiation pressure, implying an accretion-disk profile. To test whether the accretion scenario is compatible with the observed spectral profile, we computed a simplified accretion-disk profile by assuming that the accretion rate is constant and that the disk is passively heated ($T\propto r^{-1/2}$), giving a surface density $\Sigma\propto r^{-1}$, as can be derived from the equations given by \citet{Lynden-Bell_Pringle_1974}. The proportionality constant was then fitted separately for the two sides of the disk. As for all our models, the temperature profile was calculated by \textsc{ontario} and was not imposed. Thus, our accretion model is not entirely self-consistent. However, we checked that the computed temperature profile can be reasonably well approximated by $T\propto r^{-1/2}$.

The result of the fit is shown in Fig.~\ref{fig:HIFI_data_plots}c. As can be seen, an accretion disk produces too much emission from the inner regions of the disk, resulting in broad wings not seen in the data. Truncating the disk inside 30\,AU removes the wings and produces a better match to the observed profile and illustrates the paucity of C gas in the inner regions. Our conclusion is that a simplified accretion-disk model like the one we present here is not able to reproduce the observed \CII line profile.

\subsection{Constraints on the \CII radial density profile}
As a next step, we constrained the place where the carbon gas is located in the disk by using information contained in the spectrally resolved HIFI observation of the \CII 158\,$\mu m$ emission line. We produced a simplified model where each side of the disk was modelled as a series of concentric half-rings of constant density within the mid-plane of each half-ring. Since other species have been observed to show strong asymmetries between the NE and SW side of the disk \citep[see e.g.][]{Brandeker_etal_2004}, our disk models have independent density profiles for each side. If we consider $m$ rings, the density $n$ of such a profile as a function of radius $r$ and height above the mid-plane $z$  can be written
\begin{equation}
n(r,z)=\sum_{i=1}^m \Pi_i(r)n_i\exp\left({- \frac{z^2}{(\alpha r)^2}}\right)
,\end{equation}
with $\Pi_i$ a rectangular function defined by
\begin{equation*}
\Pi_i(r)= \left\{  \begin{array}{l l}
   1 & \quad \text{if $r$ within ring $i,$}\\
   0 & \quad \text{otherwise}
 \end{array} \right.
\end{equation*}
$n_i$ is the number density of carbon gas (neutral and ionised) in the mid-plane for ring $i$ and $\alpha$ denotes the opening angle of the disk. Only the $n_i$ parameters are subject to fitting, $\alpha$ is kept constant. We used $\alpha=0.2$, in line with the values derived by \citet{Brandeker_etal_2004} and \citet{Nilsson_etal_2012}. The dependence of the derived spectral profile on $\alpha$ is very weak because its principle influence is through the positioning of the emission in the beam, and optical depth effects for this moderately thick line. We chose four half-rings for each side of the disk, making a total of eight free parameters. The boundaries of the rings were chosen to be at 1, 10, 30, 150, and 300\,AU. These boundaries were chosen so that they produce spectrally resolvable contributions and at the same time trace sufficient signal to be meaningfully constrained.

From the observed line profile we can see that the contribution to emission from the region inside 1\,AU is insignificant; most of the flux would show up in the far wings of the line, where no signal is detected. Gas outside 300\,AU, on the other hand, would give rise to emission spectrally unresolved from the region 150--300\,AU, since the velocity dispersion decreases with radial distance. But we know from earlier PACS observations \citep{Brandeker_etal_2014_subm} that \CII emission was only detected in the central spaxel, so there cannot be significant emission outside 300\,AU.

\begin{figure}
\resizebox{\hsize}{!}{\includegraphics{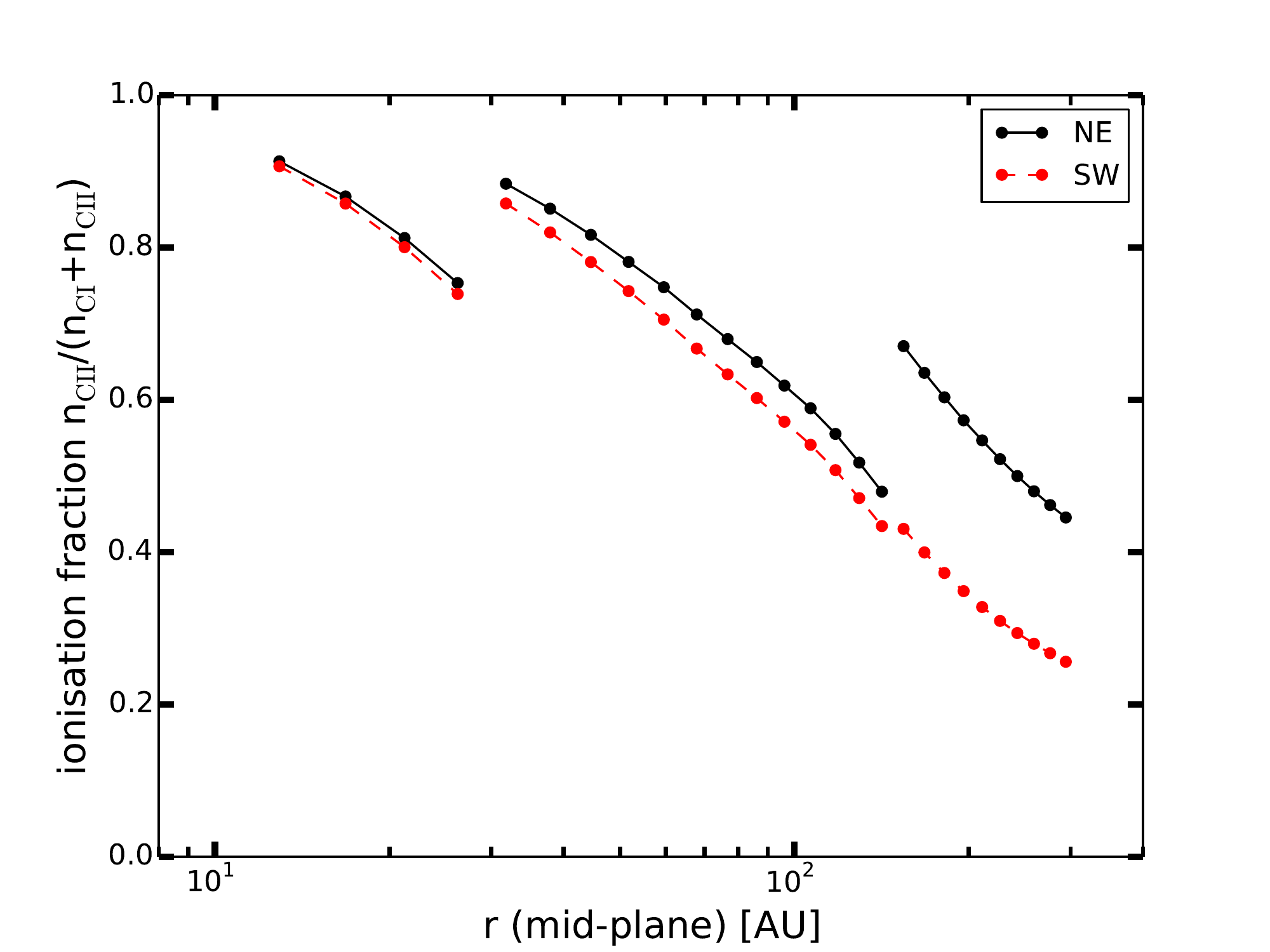}}
\caption{Ionisation fraction in the mid-plane for the best ring fit. Discontinuities indicate the borders between different rings. For this particular fit, the innermost ring (1--10\,AU) is empty, thus the ionisation fraction is not defined there. The difference in ionisation for the outermost ring (150--300\,AU) is due to the four times higher best-fit density in the SW for this ring (see Table~\ref{tab:bestfit_confidenceintervals}).}
\label{fig:ionisation_fraction}
\end{figure}

Our best fit suggests a total \ion{C}{I} mass of $7.1\times10^{-3}$~M$_\oplus$ and a total \CII mass of $5.5\times10^{-3}$~$M_\oplus$.  In Table~\ref{tab:bestfit_confidenceintervals} we present the best-fit values of the half-ring masses and densities (for total C, i.e.\ neutral and singly ionised). Errors on these values are estimated in Sect.~\ref{sec:error_bars}. Fig.~\ref{fig:HIFI_data_plots}b shows the best fit and the contribution of each ring to the fit. The ionisation fraction (in the mid-plane) is roughly 50\% in most parts of the disk, as shown in Fig.~\ref{fig:ionisation_fraction}.

If we assume that the gas mass is dominated by C and O \citep{Brandeker_etal_2014_subm}, with solar $m_{\mathrm{C}}/m_{\mathrm{O}} = 0.38$ \citep{Lodders_2003}, the implied gas-to-dust mass ratio is found to be 0.8, where the millimetre dust mass 0.06\,M$_\oplus$ \citep{Nilsson_etal_2009} was used.

\renewcommand{\arraystretch}{1.25}% this is to stretch the vertical distance between the rows
\begin{table*}
\caption{Best-fit values and central 90\% confidence intervals for the masses and densities in the mid-plane of the C gas (neutral and ionised) for each half-ring.} % title of Table
\label{tab:bestfit_confidenceintervals}
\centering % used for centering table
\begin{tabular}{c c c} % centered columns
\hline\hline % inserts double horizontal lines
Ring & NE & SW\\
& \multicolumn{2}{c}{best-fit value} \\
\hline
(AU) & \multicolumn{2}{c}{mass (M$_\oplus$)} \\
\hline % inserts single horizontal line
1--10 &  $0.0_{-0.0}^{+4.9}\times10^{-5}$ & $0.0_{-0.0}^{+5.6}\times10^{-5}$ \\ % inserting body of the table
10--30 &  $4.4_{-4.4}^{+5.6}\times 10^{-5}$ & $5.0_{-5.0}^{+6.1}\times 10^{-5}$\\
30--150 & $1.1_{-0.5}^{+1.4}\times 10^{-3}$ &  $1.4_{-0.5}^{+1.5}\times 10^{-3}$  \\
150--300 & $2.0_{-2.0}^{+6.9}\times 10^{-3}$ & $8.2_{-8.2}^{+14.1}\times 10^{-3}$ \\
total & $3.1_{-2.0}^{+7.0}\times10^{-3}$ & $9.7_{-7.6}^{+14.4}\times10^{-3}$ \\
\hline
 & \multicolumn{2}{c}{mid-plane density $n_i$ (cm$^{-3}$)} \\
 \hline
 1--10 & $0.0_{-0.0}^{+1.2}\times10^4$ & $0.0_{-0.0}^{+1.4}\times10^4$ \\
10--30 & $4.0_{-4.0}^{+5.2}\times10^2$ & $4.6_{-4.6}^{+5.6}\times10^2$ \\
30--150 & $7.6_{-3.6}^{+9.7}\times10^1$ & $1.0_{-0.3}^{+1.1}\times10^2$ \\
150--300 & $2.0_{-2.0}^{+7.1}\times10^1$ & $8.4_{-8.4}^{+14.3}\times10^1$ \\
\hline
\end{tabular}
\end{table*}

\subsection{Error analysis}\label{sec:error_bars}
To test how sensitive our fit to the data is to the estimated noise, we used a Monte Carlo approach and fitted our models to artificial data sets produced by adding synthetic noise. Since correlations between adjacent pixels are present in the data, simply adding Gaussian white noise is not appropriate; instead, we constructed correlated noise in a way described in Appendix \ref{details_error_analysis}.

As additional error sources, we included (i) the telescope pointing uncertainty by randomly varying the positions of the HIFI beams for each fit according to \textit{Herschel}'s 1$\sigma$ pointing accuracy of $\sim$2$\arcsec$ \citep{Pilbratt_etal_2010} and (ii) the uncertainty in the systematic velocity of $\beta$~Pic by randomly varying its value for each fit according to the 1$\sigma$ error of 0.2\,km\,s$^{-1}$ quoted by \citet{Brandeker_2011}.

As initial conditions for the fits, we used the best-fit values presented in Table~\ref{tab:bestfit_confidenceintervals}, except when the best-fit value was 0, as is the case for the innermost ring. To ensure that the content of the innermost ring was not underestimated, we used $2.5\times10^4$\,cm$^{-3}$ as an initial value for the mid-plane density.

\begin{figure}
\resizebox{\hsize}{!}{\includegraphics{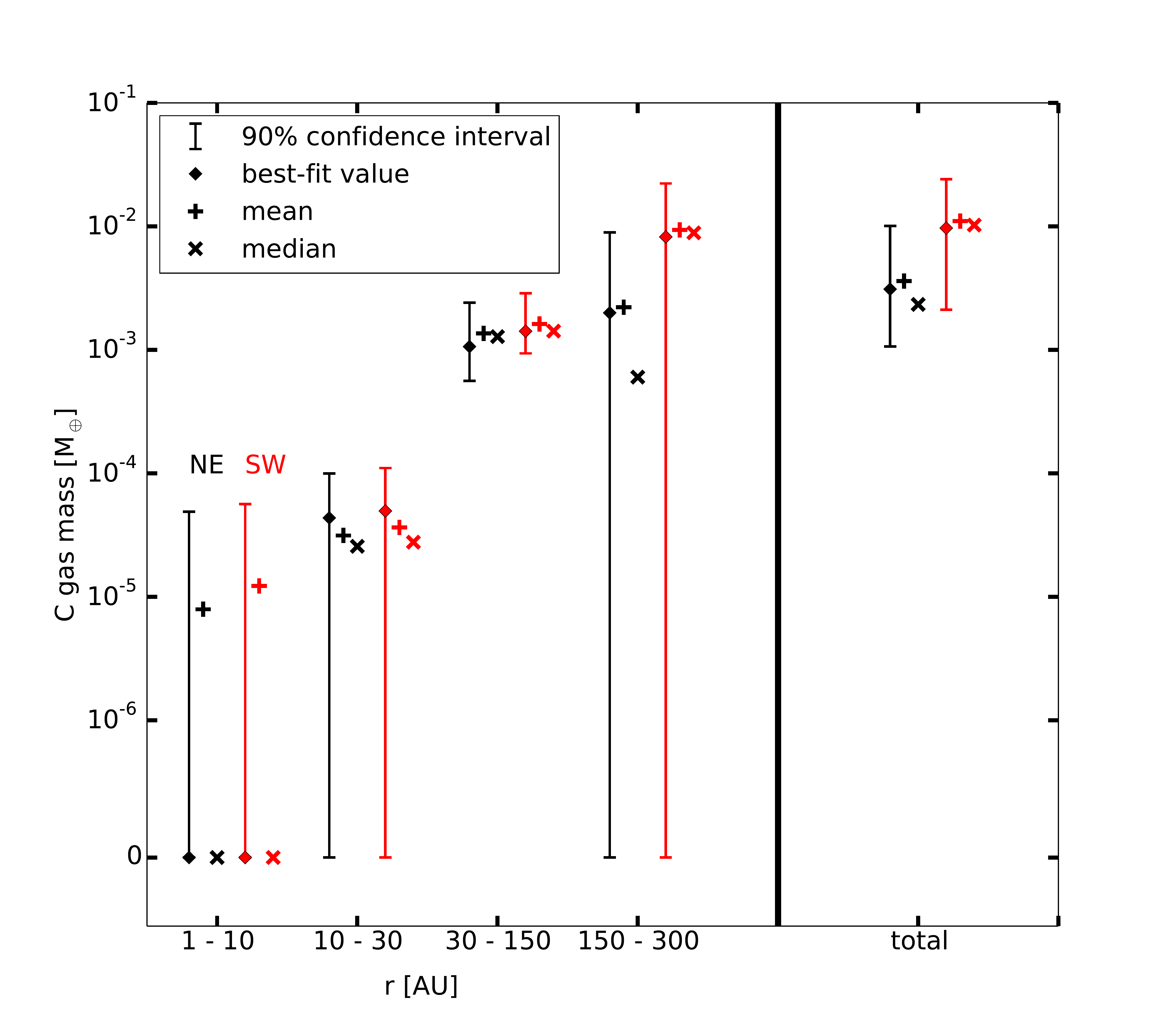}}
\caption{C gas content of half-rings and the total content of each side of the disk.}
\label{fig:error_bars}
\end{figure}

In Fig.~\ref{fig:error_bars} we present central 90\,\% confidence intervals for the mass in a given half-ring derived from a total of 837 fits. The results are summarised in Table~\ref{tab:bestfit_confidenceintervals}. The best constraints can be set on the ring spanning from 30 to 150\,AU. As can be seen from the figure, the two innermost rings contain much less gas than the outer rings. Most of the C gas mass is located beyond 30\,AU. However, this does not mean that the density in the inner rings is necessarily low. In fact, the inner rings can have much higher densities than the outer rings, since their volume is much smaller.

\subsection{Asymmetry between the NE and the SW side of the disk}\label{subsec:asymmNESW}
\begin{figure}
\resizebox{\hsize}{!}{\includegraphics{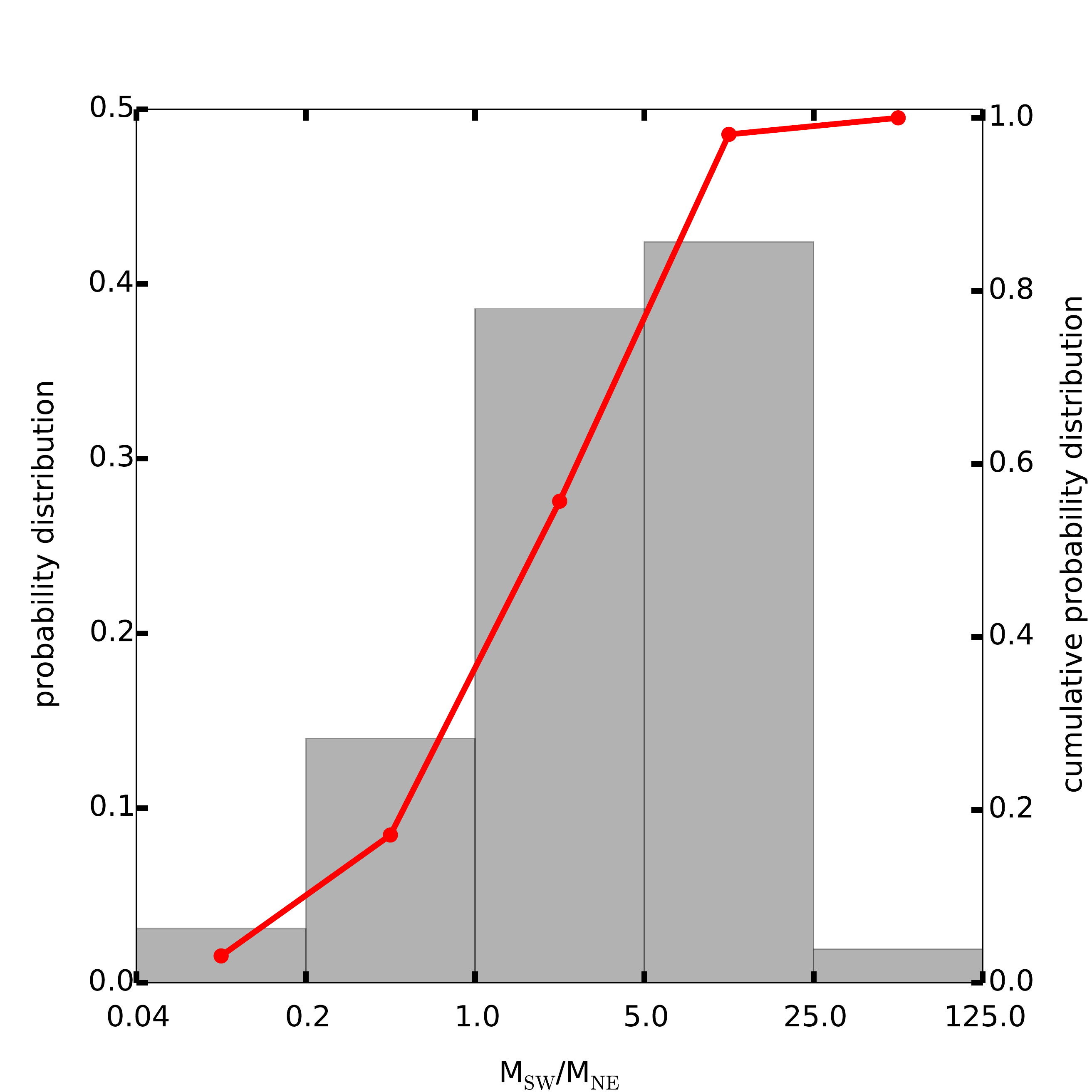}}
\caption{Histogram (probability distribution) and cumulative probability distribution (overlaid curve) of the SW-to-NE mass ratio derived from fits to artificial data sets. The edges of the bins are shown on the x-axis. 83\% of the fits have $M_\mathrm{SW}/M_\mathrm{NE}>1$.}
\label{fig:massratio_NE_SW}
\end{figure}

Especially the line profile observed with the V-polarisation beam shows more emission for velocities lower than the systematic velocity of $\beta$~Pic, indicating emission preferentially from the SW (Fig.~\ref{fig:HIFI_data_plots}). We used the fits to artificial data sets presented in Sect.~\ref{sec:error_bars} to assess the significance of this asymmetry by computing the mass ratio $M_\mathrm{SW}/M_\mathrm{NE}$ for each fit. Fig.~\ref{fig:massratio_NE_SW} shows the histogram and the cumulative probability distribution derived. 83\% of the fits have $M_\mathrm{SW}/M_\mathrm{NE}>1$. While this is certainly suggestive of a real asymmetry in the gas disk, the HIFI data are too noisy to firmly exclude a symmetric or even an NE-dominated disk. Spatially resolved data for instance\ from ALMA (Sect.~\ref{subsec:ALMA}) might be able to address this issue more directly.

\section{Discussion}\label{sec:discussion}

\subsection{Implications on the nature of the gas-producing mechanism}
Both the origin of the gas around $\beta$~Pic as well as the reason for its peculiar composition (C and O overabundance) are unknown.  If the gas is produced from the dust and is not primordial \citep{Fernandez_etal_2006}, it should be produced where most of the dust resides, that is,\ at $\sim$100\,AU. The peculiar composition can then be explained by two different pathways \citep{Xie_etal_2013}. In the first, if the gas is produced at the observed composition (e.g.\ photodesorption from C/O-rich icy grains), inward accretion is necessary and one expects to observe an accretion disk inward of 100\,AU. In the second, if the gas could be produced at normal abundances (e.g.\ collisional vaporisation) and certain elements preferentially removed by radiation pressure, no accretion disk that extends inward would be observed.

If we assume a disk surface density profile of $\Sigma \propto r^{-\gamma}$, the mass measurements in Table \ref{tab:bestfit_confidenceintervals} (taking the lower limit for 30--150\,AU and the upper limit for 10--30\,AU) require that $\gamma\leq1$. This conclusion is supported by studying the velocity profile (Fig. \ref{fig:HIFI_data_plots}c) where a disk with $\gamma = 1$ generates too much emission at high velocities. Such a shallow profile (low surface density in the inner region) is inconsistent with a standard $\alpha$-accretion disk. However, we comment that $\beta$~Pic~b, the massive giant planet observed at $r \sim 8$\,AU may perturb such a disk and give rise to a wide gap in the accretion disk. This could then explain the shallow density profile that our data suggest.

Inside of 30\,AU, noise in our data prevents us from deriving a more stringent upper limit on the gas content. More observations where the gas disk can be better resolved, for instance,\ with ALMA (Sect.\ \ref{subsec:ALMA}), are needed to constrain both the accretion disk profile and the effect of the giant planet;
this will lead to a final clarification on the gas production mechanism in the $\beta$~Pic disk.

\textit{Falling evaporation bodies} (FEBs), essentially comets evaporating in immediate vicinity to the star, have also been suggested as a possible gas source. They most likely contribute gas near the star, and it has been suggested that \ion{Ca}{II} at high latitudes is a result of gas released from inclined FEBs being pushed out by strong radiation pressure \citep{Beust_etal_2007}. However, our results show that most of the C gas mass is located beyond 30\,AU, strongly suggesting that FEBs cannot be the source of the bulk of the circumstellar gas. With FEB gas production occurring close to the star, the gas would have to be effectively transported outwards. While radiation pressure can induce an outward drift for some elements \citep[see][]{Xie_etal_2013}, the radiation pressure on carbon is negligible. In addition, carbon is believed to be the braking agent that prevents other species such as \ion{Na}{I} from quickly being blown out by radiation pressure \citep{Fernandez_etal_2006}.

\subsection{Comparison with previous observations}
Our results can be compared with previous observations by \citet{Roberge_etal_2006}. They observed \CII absorption in the far-UV and measured at total C gas column density of $\sim5\times 10^{16}$~cm$^{-2}$. Our best ring fit suggests a column density $\sim$10 times higher: $3.0_{-1.4}^{+16.1}\times 10^{17}$~cm$^{-2}$ (NE) and $5.1_{-2.0}^{+17.9}\times 10^{17}$~cm$^{-2}$ (SW), respectively, where the 90\% confidence intervals derived from our Monte Carlo simulations (Sect.~\ref{sec:error_bars}) are stated. The high upper boundaries of the confidence intervals is due to the two innermost rings, which can have very high column densities, although they contribute little to the total emission because of their small volume compared with the two outer rings (Fig.~\ref{fig:HIFI_data_plots}b) and Table~\ref{tab:bestfit_confidenceintervals}). Omitting these two rings in the calculation of the confidence intervals significantly reduces the upper boundary: we derive $1.8_{-0.6}^{+2.1}\times10^{17}$~cm$^{-2}$ (NE) and $3.7_{-1.3}^{+4.3}\times10^{17}$~cm$^{-2}$ (SW). Comparing our results with the observations of \citet{Roberge_etal_2006}, we conclude that our modelling suggests a C colum density $\sim$10 times higher than derived from the UV lines, in line with the results derived from the expected efficiency of C as a braking agent for neutral species \citep{Brandeker_2011} and recent \textit{Herschel} PACS observations of \CII\citep{Brandeker_etal_2014_subm}.

\subsection{Comparison with other gaseous debris disks around A-type stars}
There have recently been new detections, mainly thanks to {\it Herschel}'s high sensitivity, of carbon gas emission from debris disks around A-stars. \citet{Donaldson_etal_2013} observed HD\,32297 with {\it Herschel} PACS and reported a 3.7$\sigma$ detection of \CII 158\,$\mu$m emission. Assuming local thermodynamic equilibrium and optically thin emission, they derived a lower limit on the \CII disk mass of $M_{\CII}>1.7\times 10^{-4}$ M$_\oplus$. Similarly, \citet{Roberge_etal_2013} derived $M_{\CII}>2.15\times 10^{-4}$ M$_\oplus$ for the 49~Ceti debris disk. If the same processes are producing the circumstellar gas in the $\beta$~Pic disk as well as in these two disks, one would expect similar gas masses. Our results indicate $M_{\CII}\approx5.5\times10^{-3}$~M$_\oplus$ for $\beta$~Pic, well above the lower limits for HD\,32297 and 49~Cet, implying that the systems could in principle have a similar \CII gas mass. However, although the systems are potentially similar in this respect, there are other important differences. Both HD\,32297 and 49~Cet lack a detection of \ion{O}{I}, in contrast to $\beta$~Pic. HD\,32297 also shows unusually high abundance of Na, which might be due to contamination along the line of sight, however \citep{Donaldson_etal_2013}.

\subsection{Future prospects with ALMA}\label{subsec:ALMA}
Because \textsc{ontario} calculates the gas ionisation in the disk, our modelling of the \ion{C}{II} emission automatically results in a prediction of the luminosity of \ion{C}{I} at 609\,$\mu$m as well as its spatial distribution. The \textit{Atacama Large Millimeter/sub-millimeter Array} (ALMA) has the sensitivity as well as the spatial and spectral resolution to test this prediction. ALMA observations are highly desirable, since in combination with the HIFI data, they would give us a more robust estimate of the total amount of carbon in the disk and strongly constrain parameters such as the electron density and temperature, critical parameters for any gas-disk modelling (like \textsc{ontario}). ALMA observations of \ion{C}{I} would also be interesting in the context of the recent ALMA observations of CO \citep{Dent_etal_2014_subm}.

Our best-fit ring model predicts a total \ion{C}{I} 609\,$\mu$m flux of $2.8\times10^{-15}$~erg~s$^{-1}$~cm$^{-2}$. For the model derived from observations of \ion{Fe}{I} under the assumption of a well-mixed gas, this flux is reduced by a factor of $\sim$3. In this latter case, the disk extent is reduced compared with the ring fit, since the C density profile is forced to follow the observed \ion{Fe}{I} profile. The HIFI beam has a HPBW of $\sim$200\,AU and is thus not very sensitive to the additional gas present far out in the ring model. The ring model and the \ion{Fe}{I}-derived model have therefore similar projected spectra (Fig.~\ref{fig:HIFI_data_plots}a,b), but differ in flux when observed with a sufficiently large beam.

\begin{figure}
\resizebox{\hsize}{!}{\includegraphics{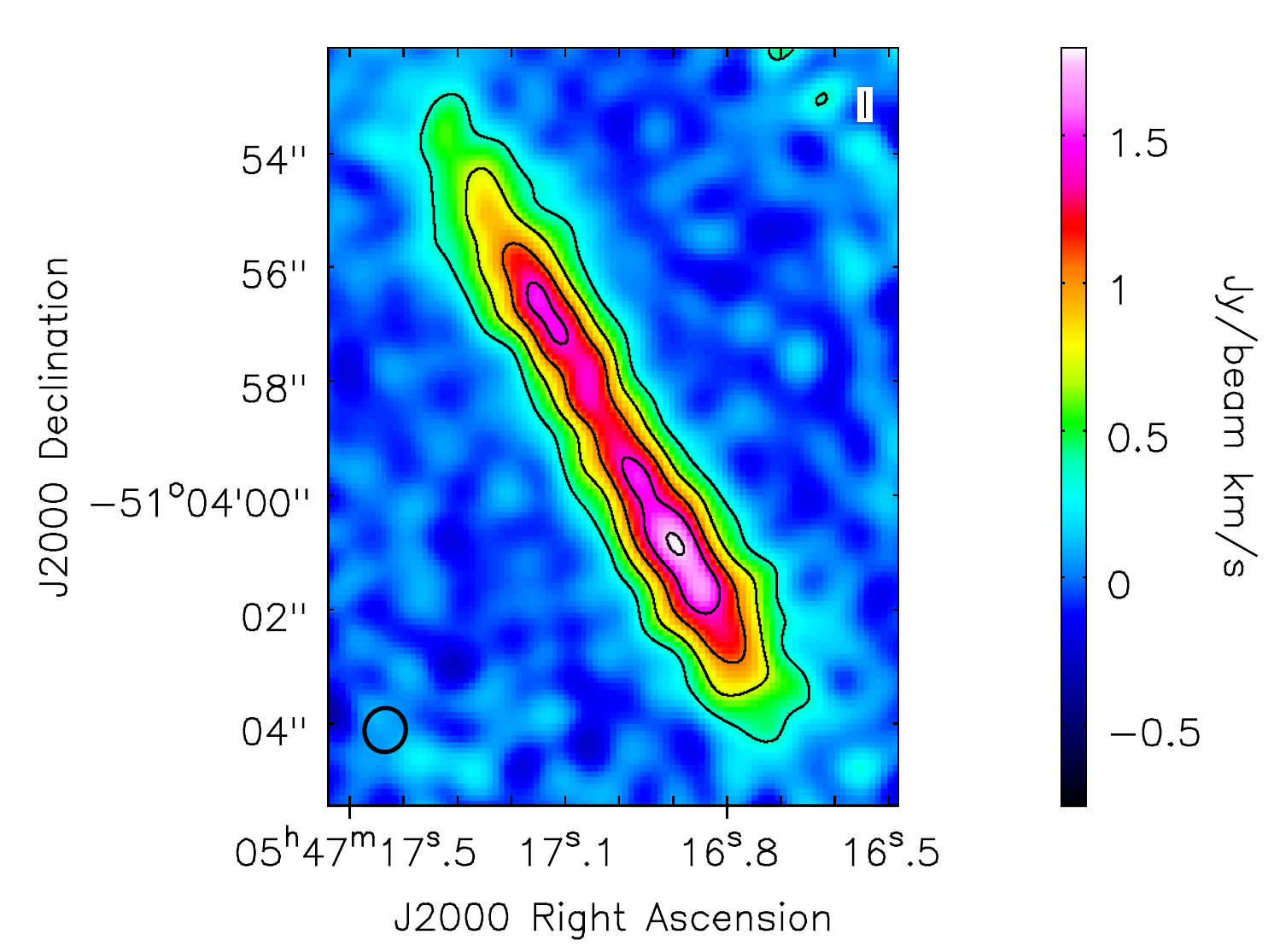}}
\caption{Simulated ALMA cycle 2 observation (using CASA) of the predicted \ion{C}{I} 609\,$\mu$m emission from the $\beta$~Pic disk, integrated over the spectral axis. With a primary beam of 12.6$\arcsec$, a mosaic of three pointings with a spacing of 6$\arcsec$ (Nyquist sampling) was sufficient to cover the disk. Contours denote 3, 6, 9, 12, and 15 times the rms noise level. The simulated observation time (on source) is 1.24\,h. We simulated the most compact cycle 2 configuration (32 12\,m-antennas, longest baseline $\sim$160\,m) and a precipitable water vapor column of 0.472\,mm (first octile). The synthesized beam (0.76$\arcsec$, corresponding to 15\,AU at the distance of $\beta$~Pic) is illustrated in the lower left corner.}
\label{fig:CI_map}
\end{figure}

We can assess the possibility of detecting \ion{C}{I} 609\,$\mu$m emission with ALMA by simulating observations using the \textit{Common Astronomy Software Applications} package \citep[CASA,][]{McMullin_etal_2007}. As input, we used our \ion{Fe}{I}-derived model (assuming well-mixed gas), which is more conservative in terms of expected flux. In Fig.~\ref{fig:CI_map} we present an emission map from simulated ALMA observations which clearly shows that ALMA is capable of detecting \ion{C}{I} 609\,$\mu$m emission from $\beta$ Pic. Thanks to the high spectral resolution of ALMA, the Keplerian velocity field can be used to distinguish emission location along the line of sight, producing a 3D map of the gas distribution. With ALMA observations, it would thus be possible for example to\ assess whether there is  an inner hole in the gas disk and distinguish between the two models presented in Fig.~\ref{fig:HIFI_data_plots}a).

\section{Summary}\label{sec:summary}
Our results are summarised as follows:
\begin{enumerate}
\item We detected and spectrally resolved \CII 158\,$\mu$m emission from the $\beta$~Pic debris disk using \textit{Herschel} HIFI.
\item We constrained the spatial distribution of the C gas in the disk by using the shape of the emission line. We estimated a total C gas mass of $1.3_{-0.5}^{+1.3}\times10^{-2}$~$M_\oplus$ (central 90\% confidence interval).
\item The observed line profile is consistent with the hypothesis of a well-mixed gas (C/Fe=300 throughout the disk).
\item Our observations argue in favour of a preferential depletion explanation for the overabundance of C and O (compared with solar abundances of metallic elements, e.g.\ Na and Fe), since the line profile is inconsistent with a simplified accretion-disk model.
\item We found that most of the gas is located beyond 30\,AU, making FEBs an unlikely source for the bulk of the $\beta$~Pic circumstellar gas.
\item The HIFI data suggest an asymmetry in the C gas disk, with the SW side being more massive than the NE side.
\item It is expected that it will be possible to detect and resolve \ion{C}{I} 609\,$\mu$m emission using ALMA.
\end{enumerate}

\appendix
\section{Details of the error analysis}\label{details_error_analysis}
\begin{figure}
\resizebox{\hsize}{!}{\includegraphics{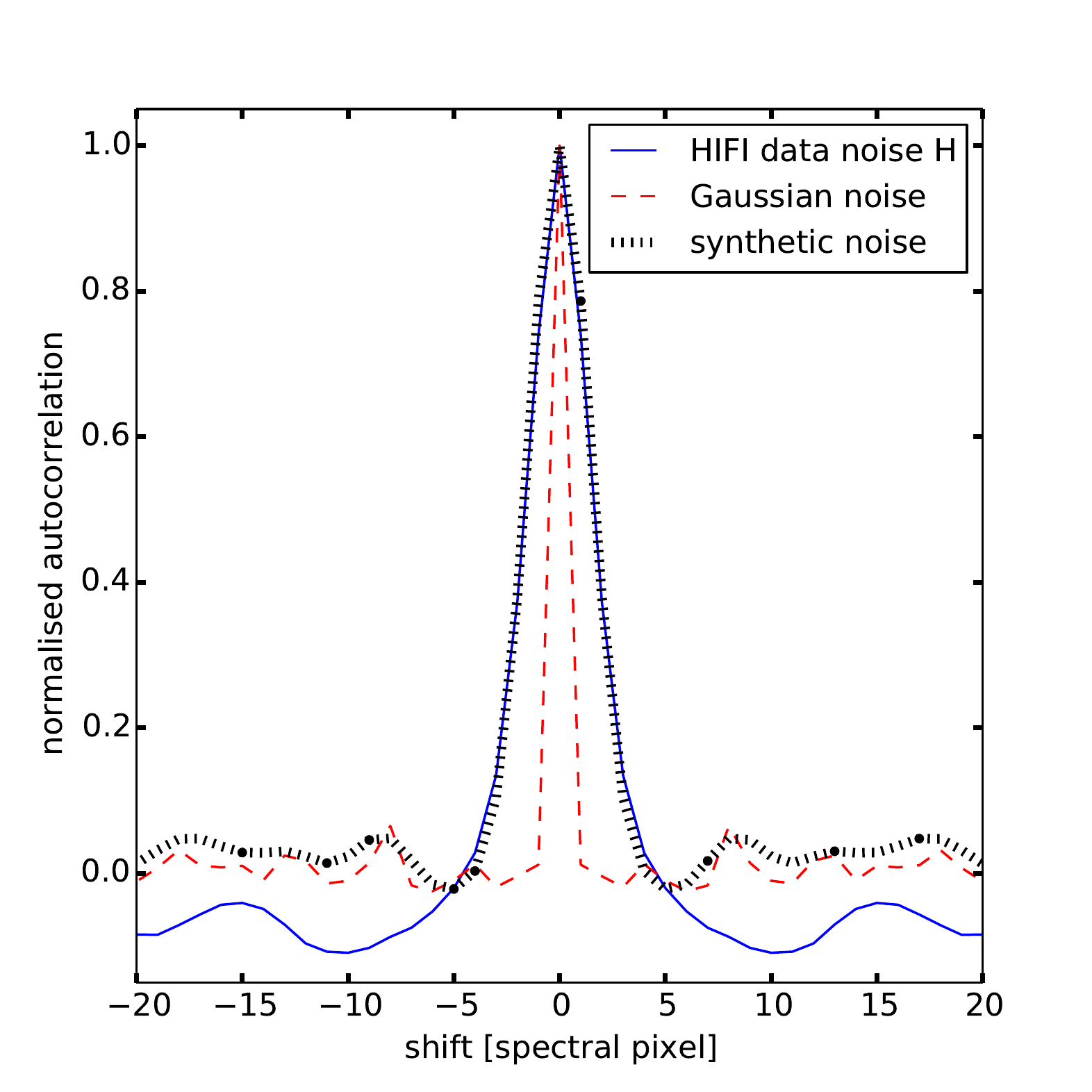}}
\caption{Autocorrelation of the HIFI noise (for the H beam; the figure for the V beam looks very similar) and of Gaussian noise for comparison. Clearly, adjacent pixels in the oversampled HIFI data used for fitting are correlated. Also shown is the autocorrelation of our synthetic noise, which well-represents the correlation pattern observed in the data. The negative autocorrelation of the data is of random nature and due to the limited number of data points. The analysed HIFI noise comes from the data interval $[-53,10]$~$\cup$~$[50,100]$ km\,s$^{-1}$, i.e.\ well outside the \CII line (Fig.~\ref{fig:HIFI_data_plots}).}
\label{autocorr}
\end{figure}

Adjacent channels in our oversampled HIFI data used for fitting are correlated in a non-negligible way. This is seen in Fig.~\ref{autocorr}, which shows the autocorrelation of the HIFI noise. When generating synthetic noise, used to add to the data for repeated fitting, we need to take the correlations into account. In other words, the generated synthetic noise needs to show the same correlation pattern (i.e.\ the same autocorrelation) as the HIFI data. One way to achieve this is by convolving Gaussian noise $g$ with an appropriate function $f$. The condition allowing us to determine $f$ reads
\begin{equation}
(f*g)\star(f*g)=a
,\end{equation}
where $x*y$ denotes the convolution, $x\star y$ the cross-correlation and $a$ is the autocorrelation of the HIFI data. This simplifies to
\begin{equation}
f\star f=a
\end{equation}
using properties of the cross-correlation and the convolution and the fact that $g$ is Gaussian noise. It is then straightforward to show that $f$ is given by
\begin{equation}
f=\mathscr{F}^{-1}\left(\sqrt{\mathscr{F}(a)}\right)
,\end{equation}
with $\mathscr{F}(x)$ the Fourier transform. Modelling $a$ as a Gaussian with standard deviation $\sigma_a$, fitted to the measured autocorrelation shown in Fig.~\ref{autocorr}, we find that $f$ is a Gaussian as well, with standard deviation
\begin{equation}
\sigma_f=\frac{\sigma_a}{\sqrt{2}}
.\end{equation}
Fig.~\ref{autocorr} shows that synthetic noise generated in this way well represents the correlated noise observed in the HIFI data. Note that the negative autocorrelation of the HIFI data (H polarisation) for shifts larger than five pixels is of random nature due to the limited number of data points available. Indeed, the autocorrelation of the V-polarisation data is negative as well for shifts larger than five pixels, but becomes positive for shifts larger than 14 pixels. Additional evidence for a random nature of the feature comes from the synthetic noise, which shows similar oscillations (Fig.~\ref{autocorr}). If constructed from a sufficiently large data set, these oscillations disappear.

\begin{acknowledgements}
We thank the anonymous referee for improving this paper with useful comments and constructive criticism. We would also like to thank Inga Kamp, Sarah Jabbari and Eric Stempels for useful discussions. This research has made use of the SIMBAD database (operated at CDS, Strasbourg, France), the NIST Atomic Spectra Database and NASA's Astrophysics Data System. AB was supported by the \textit{Swedish National Space Board} (contract 113/10:3).
\end{acknowledgements}

\bibpunct{(}{)}{;}{a}{}{,} % to follow the A&A style


\begin{thebibliography}{54}
\expandafter\ifx\csname natexlab\endcsname\relax\def\natexlab#1{#1}\fi

\bibitem[{{Aumann}(1985)}]{Aumann_1985}
{Aumann}, H.~H. 1985, \pasp, 97, 885

\bibitem[{{Aumann} {et~al.}(1984){Aumann}, {Beichman}, {Gillett}, {de Jong},
  {Houck}, {Low}, {Neugebauer}, {Walker}, \& {Wesselius}}]{Aumann_etal_1984}
{Aumann}, H.~H., {Beichman}, C.~A., {Gillett}, F.~C., {et~al.} 1984, \apjl,
  278, L23

\bibitem[{{Backman} \& {Paresce}(1993)}]{Backman_etal_1993}
{Backman}, D.~E. \& {Paresce}, F. 1993, in Protostars and Planets III, ed.
  E.~H. {Levy} \& J.~I. {Lunine}, 1253--1304

\bibitem[{{Beust} \& {Valiron}(2007)}]{Beust_etal_2007}
{Beust}, H. \& {Valiron}, P. 2007, \aap, 466, 201

\bibitem[{{Binks} \& {Jeffries}(2014)}]{Binks_Jeffries_2014}
{Binks}, A.~S. \& {Jeffries}, R.~D. 2014, \mnras, 438, L11

\bibitem[{{Brandeker}(2011)}]{Brandeker_2011}
{Brandeker}, A. 2011, \apj, 729, 122

\bibitem[{{Brandeker} {et~al.}(2004){Brandeker}, {Liseau}, {Olofsson}, \&
  {Fridlund}}]{Brandeker_etal_2004}
{Brandeker}, A., {Liseau}, R., {Olofsson}, G., \& {Fridlund}, M. 2004, \aap,
  413, 681

\bibitem[{{Brandeker} {et~al.}(2014){Brandeker}, {Olofsson}, {Vandenbussche},
  {Acke}, {Barlow}, {Blommaert}, {Cohen}, \& {Dent}}]{Brandeker_etal_2014_subm}
{Brandeker}, A., {Olofsson}, G., {Vandenbussche}, B., {et~al.} 2014, subm. to
  \aap

\bibitem[{{Chen} {et~al.}(2007){Chen}, {Li}, {Bohac}, {Kim}, {Watson}, {van
  Cleve}, {Houck}, {Stapelfeldt}, {Werner}, {Rieke}, {Su}, {Marengo},
  {Backman}, {Beichman}, \& {Fazio}}]{Chen_etal_2007}
{Chen}, C.~H., {Li}, A., {Bohac}, C., {et~al.} 2007, \apj, 666, 466

\bibitem[{{Crawford} {et~al.}(1994){Crawford}, {Spyromilio}, {Barlow}, {Diego},
  \& {Lagrange}}]{Crawford_etal_1994}
{Crawford}, I.~A., {Spyromilio}, J., {Barlow}, M.~J., {Diego}, F., \&
  {Lagrange}, A.~M. 1994, \mnras, 266, L65

\bibitem[{{Crifo} {et~al.}(1997){Crifo}, {Vidal-Madjar}, {Lallement}, {Ferlet},
  \& {Gerbaldi}}]{Crifo_etal_1997}
{Crifo}, F., {Vidal-Madjar}, A., {Lallement}, R., {Ferlet}, R., \& {Gerbaldi},
  M. 1997, \aap, 320, L29

\bibitem[{{Czechowski} \& {Mann}(2007)}]{Czechowski_etal_2007}
{Czechowski}, A. \& {Mann}, I. 2007, \apj, 660, 1541

\bibitem[{{de Graauw} {et~al.}(2010){de Graauw}, {Helmich}, {Phillips},
  {Stutzki}, {Caux}, {Whyborn}, {Dieleman}, {Roelfsema}, {Aarts}, {Assendorp},
  {Bachiller}, {Baechtold}, {Barcia}, {Beintema}, {Belitsky}, {Benz}, {Bieber},
  {Boogert}, {Borys}, {Bumble}, {Ca{\"i}s}, {Caris}, {Cerulli-Irelli},
  {Chattopadhyay}, {Cherednichenko}, {Ciechanowicz}, {Coeur-Joly}, {Comito},
  {Cros}, {de Jonge}, {de Lange}, {Delforges}, {Delorme}, {den Boggende},
  {Desbat}, {Diez-Gonz{\'a}lez}, {di Giorgio}, {Dubbeldam}, {Edwards},
  {Eggens}, {Erickson}, {Evers}, {Fich}, {Finn}, {Franke}, {Gaier}, {Gal},
  {Gao}, {Gallego}, {Gauffre}, {Gill}, {Glenz}, {Golstein}, {Goulooze},
  {Gunsing}, {G{\"u}sten}, {Hartogh}, {Hatch}, {Higgins}, {Honingh}, {Huisman},
  {Jackson}, {Jacobs}, {Jacobs}, {Jarchow}, {Javadi}, {Jellema}, {Justen},
  {Karpov}, {Kasemann}, {Kawamura}, {Keizer}, {Kester}, {Klapwijk}, {Klein},
  {Kollberg}, {Kooi}, {Kooiman}, {Kopf}, {Krause}, {Krieg}, {Kramer},
  {Kruizenga}, {Kuhn}, {Laauwen}, {Lai}, {Larsson}, {Leduc}, {Leinz}, {Lin},
  {Liseau}, {Liu}, {Loose}, {L{\'o}pez-Fernandez}, {Lord}, {Luinge}, {Marston},
  {Mart{\'{\i}}n-Pintado}, {Maestrini}, {Maiwald}, {McCoey}, {Mehdi}, {Megej},
  {Melchior}, {Meinsma}, {Merkel}, {Michalska}, {Monstein}, {Moratschke},
  {Morris}, {Muller}, {Murphy}, {Naber}, {Natale}, {Nowosielski}, {Nuzzolo},
  {Olberg}, {Olbrich}, {Orfei}, {Orleanski}, {Ossenkopf}, {Peacock}, {Pearson},
  {Peron}, {Phillip-May}, {Piazzo}, {Planesas}, {Rataj}, {Ravera}, {Risacher},
  {Salez}, {Samoska}, {Saraceno}, {Schieder}, {Schlecht}, {Schl{\"o}der},
  {Schm{\"u}lling}, {Schultz}, {Schuster}, {Siebertz}, {Smit}, {Szczerba},
  {Shipman}, {Steinmetz}, {Stern}, {Stokroos}, {Teipen}, {Teyssier}, {Tils},
  {Trappe}, {van Baaren}, {van Leeuwen}, {van de Stadt}, {Visser}, {Wildeman},
  {Wafelbakker}, {Ward}, {Wesselius}, {Wild}, {Wulff}, {Wunsch}, {Tielens},
  {Zaal}, {Zirath}, {Zmuidzinas}, \& {Zwart}}]{deGraauw_etal_2010}
{de Graauw}, T., {Helmich}, F.~P., {Phillips}, T.~G., {et~al.} 2010, \aap, 518,
  L6

\bibitem[{{Dent} {et~al.}(2014){Dent}, {Wyatt}, {Roberge}, {Augereau},
  {Casassus}, {Corder}, {Greaves}, {de Gregorio}, {Hales}, {Lagrange},
  {Matthews}, \& {Wilner}}]{Dent_etal_2014_subm}
{Dent}, W.~R.~F., {Wyatt}, M.~C., {Roberge}, A., {et~al.} 2014, subm.\ to
  Science

\bibitem[{{Donaldson} {et~al.}(2013){Donaldson}, {Lebreton}, {Roberge},
  {Augereau}, \& {Krivov}}]{Donaldson_etal_2013}
{Donaldson}, J.~K., {Lebreton}, J., {Roberge}, A., {Augereau}, J.-C., \&
  {Krivov}, A.~V. 2013, \apj, 772, 17

\bibitem[{{Fern{\'a}ndez} {et~al.}(2006){Fern{\'a}ndez}, {Brandeker}, \&
  {Wu}}]{Fernandez_etal_2006}
{Fern{\'a}ndez}, R., {Brandeker}, A., \& {Wu}, Y. 2006, \apj, 643, 509

\bibitem[{{Gorti} \& {Hollenbach}(2004)}]{Gorti_et_al_2004}
{Gorti}, U. \& {Hollenbach}, D. 2004, \apj, 613, 424

\bibitem[{{Gray} {et~al.}(2006){Gray}, {Corbally}, {Garrison}, {McFadden},
  {Bubar}, {McGahee}, {O'Donoghue}, \& {Knox}}]{Gray_etal_2006}
{Gray}, R.~O., {Corbally}, C.~J., {Garrison}, R.~F., {et~al.} 2006, \aj, 132,
  161

\bibitem[{{Haisch} {et~al.}(2001){Haisch}, {Lada}, \&
  {Lada}}]{Haisch_etal_2001}
{Haisch}, Jr., K.~E., {Lada}, E.~A., \& {Lada}, C.~J. 2001, \apjl, 553, L153

\bibitem[{{Hillenbrand}(2008)}]{Hillenbrand_2008}
{Hillenbrand}, L.~A. 2008, Physica Scripta Volume T, 130, 014024

\bibitem[{{Hobbs} {et~al.}(1985){Hobbs}, {Vidal-Madjar}, {Ferlet}, {Albert}, \&
  {Gry}}]{Hobbs_etal_1985}
{Hobbs}, L.~M., {Vidal-Madjar}, A., {Ferlet}, R., {Albert}, C.~E., \& {Gry}, C.
  1985, \apjl, 293, L29

\bibitem[{{Jayawardhana} {et~al.}(2006){Jayawardhana}, {Coffey}, {Scholz},
  {Brandeker}, \& {van Kerkwijk}}]{Jayawardhana_etal_2006}
{Jayawardhana}, R., {Coffey}, J., {Scholz}, A., {Brandeker}, A., \& {van
  Kerkwijk}, M.~H. 2006, \apj, 648, 1206

\bibitem[{{Kenyon} \& {Bromley}(2006)}]{Kenyon_etal_2006}
{Kenyon}, S.~J. \& {Bromley}, B.~C. 2006, \aj, 131, 1837

\bibitem[{{Kondo} \& {Bruhweiler}(1985)}]{Kondo_etal_1985}
{Kondo}, Y. \& {Bruhweiler}, F.~C. 1985, \apjl, 291, L1

\bibitem[{Kramida {et~al.}(2013)Kramida, {Yu.~Ralchenko}, Reader, \& {and NIST
  ASD Team}}]{NIST_ASD}
Kramida, A., {Yu.~Ralchenko}, Reader, J., \& {and NIST ASD Team}. 2013, {NIST
  Atomic Spectra Database (ver. 5.1), [Online]. Available:
  {\tt{http://physics.nist.gov/asd}} [2013, October 23]. National Institute of
  Standards and Technology, Gaithersburg, MD.}

\bibitem[{{Lagrange} {et~al.}(2012){Lagrange}, {Boccaletti}, {Milli},
  {Chauvin}, {Bonnefoy}, {Mouillet}, {Augereau}, {Girard}, {Lacour}, \&
  {Apai}}]{Lagrange_etal_2012}
{Lagrange}, A.-M., {Boccaletti}, A., {Milli}, J., {et~al.} 2012, \aap, 542, A40

\bibitem[{{Lagrange} {et~al.}(2010){Lagrange}, {Bonnefoy}, {Chauvin}, {Apai},
  {Ehrenreich}, {Boccaletti}, {Gratadour}, {Rouan}, {Mouillet}, {Lacour}, \&
  {Kasper}}]{Lagrange_etal_2010}
{Lagrange}, A.-M., {Bonnefoy}, M., {Chauvin}, G., {et~al.} 2010, Science, 329,
  57

\bibitem[{{Lodders}(2003)}]{Lodders_2003}
{Lodders}, K. 2003, \apj, 591, 1220

\bibitem[{{Lynden-Bell} \& {Pringle}(1974)}]{Lynden-Bell_Pringle_1974}
{Lynden-Bell}, D. \& {Pringle}, J.~E. 1974, \mnras, 168, 603

\bibitem[{{Mamajek}(2009)}]{Mamajek_2009}
{Mamajek}, E.~E. 2009, in American Institute of Physics Conference Series, Vol.
  1158, American Institute of Physics Conference Series, ed. T.~{Usuda},
  M.~{Tamura}, \& M.~{Ishii}, 3--10

\bibitem[{{McMullin} {et~al.}(2007){McMullin}, {Waters}, {Schiebel}, {Young},
  \& {Golap}}]{McMullin_etal_2007}
{McMullin}, J.~P., {Waters}, B., {Schiebel}, D., {Young}, W., \& {Golap}, K.
  2007, in Astronomical Society of the Pacific Conference Series, Vol. 376,
  Astronomical Data Analysis Software and Systems XVI, ed. R.~A. {Shaw},
  F.~{Hill}, \& D.~J. {Bell}, 127

\bibitem[{{Mo{\'o}r} {et~al.}(2011){Mo{\'o}r}, {{\'A}brah{\'a}m}, {Juh{\'a}sz},
  {Kiss}, {Pascucci}, {K{\'o}sp{\'a}l}, {Apai}, {Henning}, {Csengeri}, \&
  {Grady}}]{Moor_etal_2011}
{Mo{\'o}r}, A., {{\'A}brah{\'a}m}, P., {Juh{\'a}sz}, A., {et~al.} 2011, \apjl,
  740, L7

\bibitem[{{Nilsson} {et~al.}(2012){Nilsson}, {Brandeker}, {Olofsson}, {Fathi},
  {Th{\'e}bault}, \& {Liseau}}]{Nilsson_etal_2012}
{Nilsson}, R., {Brandeker}, A., {Olofsson}, G., {et~al.} 2012, \aap, 544, A134

\bibitem[{{Nilsson} {et~al.}(2009){Nilsson}, {Liseau}, {Brandeker}, {Olofsson},
  {Risacher}, {Fridlund}, \& {Pilbratt}}]{Nilsson_etal_2009}
{Nilsson}, R., {Liseau}, R., {Brandeker}, A., {et~al.} 2009, \aap, 508, 1057

\bibitem[{{Olofsson} {et~al.}(2001){Olofsson}, {Liseau}, \&
  {Brandeker}}]{Olofsson_etal_2001}
{Olofsson}, G., {Liseau}, R., \& {Brandeker}, A. 2001, \apjl, 563, L77

\bibitem[{{Ott}(2010)}]{Ott_2010}
{Ott}, S. 2010, in Astronomical Society of the Pacific Conference Series, Vol.
  434, Astronomical Data Analysis Software and Systems XIX, ed. Y.~{Mizumoto},
  K.-I. {Morita}, \& M.~{Ohishi}, 139

\bibitem[{{Pilbratt} {et~al.}(2010){Pilbratt}, {Riedinger}, {Passvogel},
  {Crone}, {Doyle}, {Gageur}, {Heras}, {Jewell}, {Metcalfe}, {Ott}, \&
  {Schmidt}}]{Pilbratt_etal_2010}
{Pilbratt}, G.~L., {Riedinger}, J.~R., {Passvogel}, T., {et~al.} 2010, \aap,
  518, L1

\bibitem[{{Redfield}(2007)}]{Redfield_2007}
{Redfield}, S. 2007, \apjl, 656, L97

\bibitem[{{Riviere-Marichalar} {et~al.}(2012){Riviere-Marichalar}, {Barrado},
  {Augereau}, {Thi}, {Roberge}, {Eiroa}, {Montesinos}, {Meeus}, {Howard},
  {Sandell}, {Duch{\^e}ne}, {Dent}, {Lebreton}, {Mendigut{\'{\i}}a},
  {Hu{\'e}lamo}, {M{\'e}nard}, \& {Pinte}}]{Riviere-Marichalar_etal_2012}
{Riviere-Marichalar}, P., {Barrado}, D., {Augereau}, J.-C., {et~al.} 2012,
  \aap, 546, L8

\bibitem[{{Roberge} {et~al.}(2006){Roberge}, {Feldman}, {Weinberger},
  {Deleuil}, \& {Bouret}}]{Roberge_etal_2006}
{Roberge}, A., {Feldman}, P.~D., {Weinberger}, A.~J., {Deleuil}, M., \&
  {Bouret}, J.-C. 2006, \nat, 441, 724

\bibitem[{{Roberge} {et~al.}(2013){Roberge}, {Kamp}, {Montesinos}, {Dent},
  {Meeus}, {Donaldson}, {Olofsson}, {Mo{\'o}r}, {Augereau}, {Howard}, {Eiroa},
  {Thi}, {Ardila}, {Sandell}, \& {Woitke}}]{Roberge_etal_2013}
{Roberge}, A., {Kamp}, I., {Montesinos}, B., {et~al.} 2013, \apj, 771, 69

\bibitem[{{Roelfsema} {et~al.}(2012){Roelfsema}, {Helmich}, {Teyssier},
  {Ossenkopf}, {Morris}, {Olberg}, {Shipman}, {Risacher}, {Akyilmaz},
  {Assendorp}, {Avruch}, {Beintema}, {Biver}, {Boogert}, {Borys}, {Braine},
  {Caris}, {Caux}, {Cernicharo}, {Coeur-Joly}, {Comito}, {de Lange},
  {Delforge}, {Dieleman}, {Dubbeldam}, {de Graauw}, {Edwards}, {Fich},
  {Flederus}, {Gal}, {di Giorgio}, {Herpin}, {Higgins}, {Hoac}, {Huisman},
  {Jarchow}, {Jellema}, {de Jonge}, {Kester}, {Klein}, {Kooi}, {Kramer},
  {Laauwen}, {Larsson}, {Leinz}, {Lord}, {Lorenzani}, {Luinge}, {Marston},
  {Mart{\'{\i}}n-Pintado}, {McCoey}, {Melchior}, {Michalska}, {Moreno},
  {M{\"u}ller}, {Nowosielski}, {Okada}, {Orlea{\'n}ski}, {Phillips}, {Pearson},
  {Rabois}, {Ravera}, {Rector}, {Rengel}, {Sagawa}, {Salomons},
  {S{\'a}nchez-Su{\'a}rez}, {Schieder}, {Schl{\"o}der}, {Schm{\"u}lling},
  {Soldati}, {Stutzki}, {Thomas}, {Tielens}, {Vastel}, {Wildeman}, {Xie},
  {Xilouris}, {Wafelbakker}, {Whyborn}, {Zaal}, {Bell}, {Bjerkeli}, {De Beck},
  {Cavali{\'e}}, {Crockett}, {Hily-Blant}, {Kama}, {Kaminski}, {Lefl{\'o}ch},
  {Lombaert}, {de Luca}, {Makai}, {Marseille}, {Nagy}, {Pacheco}, {van der
  Wiel}, {Wang}, \& {Y{\i}ld{\i}z}}]{Roelfsema_etal_2012}
{Roelfsema}, P.~R., {Helmich}, F.~P., {Teyssier}, D., {et~al.} 2012, \aap, 537,
  A17

\bibitem[{{Rybicki} \& {Lightman}(2004)}]{Rybicki_Lightman_2004}
{Rybicki}, G.~B. \& {Lightman}, A.~P. 2004, {Radiative processes in
  astrophysics} (Wiley-VCH)

\bibitem[{{Slettebak}(1975)}]{Slettebak_1975}
{Slettebak}, A. 1975, \apj, 197, 137

\bibitem[{{Slettebak} \& {Carpenter}(1983)}]{Slettebak_etal_1983}
{Slettebak}, A. \& {Carpenter}, K.~G. 1983, \apjs, 53, 869

\bibitem[{{Smith} \& {Terrile}(1984)}]{Smith_etal_1984}
{Smith}, B.~A. \& {Terrile}, R.~J. 1984, Science, 226, 1421

\bibitem[{{Tielens} \& {Hollenbach}(1985)}]{Tielens_etal_1985}
{Tielens}, A.~G.~G.~M. \& {Hollenbach}, D. 1985, \apj, 291, 747

\bibitem[{{van Leeuwen}(2007)}]{vanLeeuwen_2007}
{van Leeuwen}, F. 2007, \aap, 474, 653

\bibitem[{{Wilner} {et~al.}(2011){Wilner}, {Andrews}, \&
  {Hughes}}]{Wilner_etal_2011}
{Wilner}, D.~J., {Andrews}, S.~M., \& {Hughes}, A.~M. 2011, \apjl, 727, L42

\bibitem[{{Wyatt}(2008)}]{Wyatt_2008}
{Wyatt}, M.~C. 2008, \araa, 46, 339

\bibitem[{{Xie} {et~al.}(2013){Xie}, {Brandeker}, \& {Wu}}]{Xie_etal_2013}
{Xie}, J.-W., {Brandeker}, A., \& {Wu}, Y. 2013, \apj, 762, 114

\bibitem[{{Zagorovsky} {et~al.}(2010){Zagorovsky}, {Brandeker}, \&
  {Wu}}]{Zagorovsky_etal_2010}
{Zagorovsky}, K., {Brandeker}, A., \& {Wu}, Y. 2010, \apj, 720, 923

\bibitem[{{Zuckerman} \& {Song}(2012)}]{Zuckerman_Song_2012}
{Zuckerman}, B. \& {Song}, I. 2012, \apj, 758, 77

\bibitem[{{Zuckerman} {et~al.}(2001){Zuckerman}, {Song}, {Bessell}, \&
  {Webb}}]{Zuckerman_etal_2001}
{Zuckerman}, B., {Song}, I., {Bessell}, M.~S., \& {Webb}, R.~A. 2001, \apjl,
  562, L87

\end{thebibliography}
\end{document}